\documentclass[12pt]{article}

\usepackage{graphics,epsfig}
\begin{document}
\title{Stability of leadership in bottom-up hierarchical organizations}
\author{Serge Galam\thanks{E-mail:  galam@shs.polytechnique.fr}\\
Centre de Recherche en \'Epist\'emologie Appliqu\'ee,\\
CREA-\'Ecole Polytechnique, CNRS UMR 7656,\\
1, rue Descartes, F-75005 Paris, France}
\date{(April 8, 2005)}
\maketitle
%%%%%%%%%%%%%%%%%%%%%
\begin{abstract}

The stability of a leadership against a growing internal opposition is studied in bottom-up hierarchical organizations. Using a very simple model with bottom-up majority rule voting,  the dynamics of power distribution at the various hierarchical levels is  calculated within a probabilistic framework. Given a leadership at the top, the opposition weight from the hierarchy bottom is shown to fall off quickly while climbing up the hierarchy. It reaches zero after only a few hierarchical levels. Indeed the voting process is found to obey a threshold dynamics with a deterministic top outcome. Accordingly the leadership  may stay stable against very large amplitude increases in the opposition at the bottom level. An opposition  can thus grow steadily from few percent up to seventy seven percent with not one a single change at the elected top level. However and in contrast, from one election to another, in the vicinity of the threshold, less than a one percent additional shift at the bottom level can drive a drastic and brutal change at the top. The opposition topples the current leadership at once. In addition to analytical formulas, results from a large scale simulation are presented. The results may shed a new light on management architectures as well as on alert systems. They could also provide some different insight  on last century's  Eastern European communist collapse.\\ \\

Keywords: Local majority rules, bottom-up hierarchies, statistical physics

\end{abstract}
%%%%%%%%%%%%%%%%%%%%%%%
\section{Introduction}

Many human systems are organized within pyramidal hierarchies including large corporations, universities, armies, trade unions or political parties
\cite{general}. While many structural organizational frames are possible in practice,
two extreme and opposite cases can be formally singled out as follows. On the one side stands 
the dictatorship pyramid
initiated
at the top of the hierarchy. Decisions go down towards the base level by level. It is totally directed and deterministic with a top-down splitting
dynamics. On the other side is the perfectly democratic pyramid built up also level by level, but now
from the bottom upwards to  the top. The dynamics is bottom-up aggregating with the use of local majority rules. In this paper we  investigate the second case. For a review on voting systems see
\cite{review-voting}.

More precisely we study a generic model, which exemplifies a dynamical paradox induced by the use of bottom-up majority rules. Considering  hierarchical organizations, local democratic votes are thus shown to produce a self-neutralization of a growing internal opposition while climbing up the hierarchy from  bottom to the top.
Within a very simple probabilistic frame we show how a top leadership can hamper democratically very large amplitude changes at the hierarchy bottom from a wide support to a huge opposition. Although  we are illustrating the dynamical paradox in terms of voting process, the model is quite generic and may apply to many different situations where bottom-up majority rules aggregation are used. The frame may be a political group, a firm, a social organization.

Using tools and concepts from statistical physics, the voting process is found to obey a threshold dynamics with a deterministic top outcome \cite{grano,schelling,lemonde}. Accordingly as long as the opposition weight is below some critical threshold at the bottom, it is found to fall off quickly to reach zero after only a few hierarchical levels. It thus makes the current leadership very stable. 
In some cases, the opposition  can grow steadily from few percents up to seventy seven percents with not a single change at the top level. However and in contrast to such a stability,  in the vicinity of the threshold, from one election to another less than a one percent additional shift at the bottom level in favor of the opposition can drive a drastic and brutal change at the top. Seventy percents increase in bottom opposition has no effect on the elected organization president while less than a one percent change topples it democratically at once.

In this paper,  we consider the situation  \cite{math-psy} where members of an 
organization may support either one of two policies denoted respectively A and B. At the bottom, once each member has decided on which policy, A or B it supports, elections are set via the formation of small groups of people whose members are chosen randomly.  It means that members of a voting group are not asked about their choice before the voting. Then, each group elects a representative either A
or B according to the majority of its members. All these elected people constitute  the first hierarchical level of the organization. The voting process is then repeated now from the first level to build the second one. At each step up, stands  a smaller number of representatives with a single one at the last level, the president.  One illustration of the process is shown in Figure (1) in the case of groups of size 3.

\begin{figure}
\begin{center}
\centerline{\epsfbox{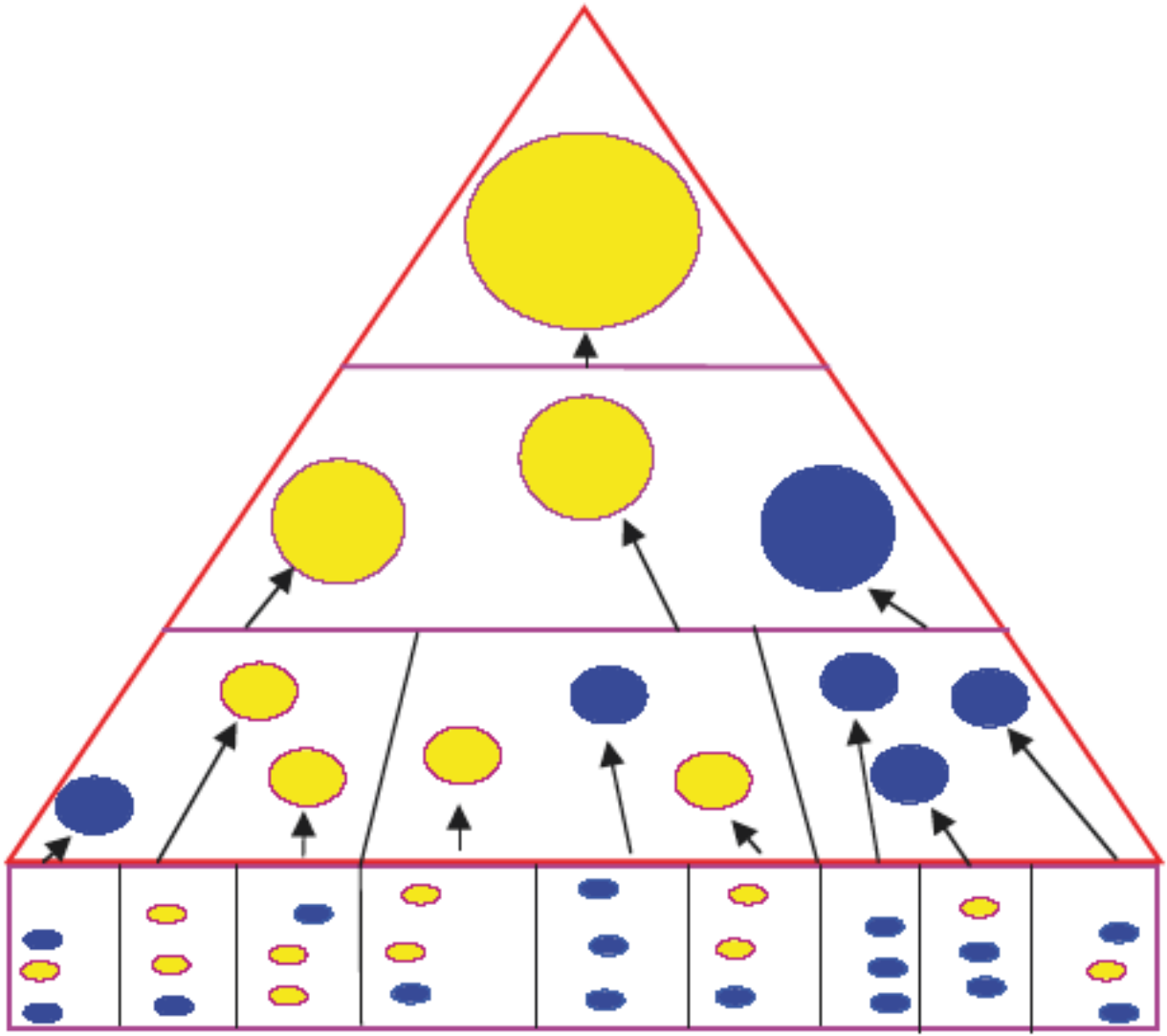}}
\caption{A two-level bottom-up hierarchy with groups of size 3}
\end{center}
\end{figure}

Accordingly, having an initial support above the critical threshold guarantees to win
the top level provided there exists some minimum number of hierarchical levels. The value of the critical threshold to power is a function of  the size of  the voting groups. For all groups of odd sizes $p_{c}=\frac{1}{2}$. However in case of 
groups of even sizes, $p_{c}neq\frac{1}{2}$. This asymmetry arises from the existence of well kwon inertia of power and status quo \cite{status}. Within our model, in the case of a local tie in an even size  group, it is the former  ruler who is reelected. 

For groups of size 4  it makes the threshold to jump at $p_{c}=0.77$ for the opposition. It means that if B is already in power, the opposition A needs to win more than $0.77 \%$ support at the bottom level to reach the top of the hierarchy. Simultaneously, the threshold to stay in power shrinks to  $0.23 \%$ to the ruling party. In addition, in cases of even size groups, the mechanism of 
self-elimination is even
faster than in the case of odd  groups. For instance, a proportion of
$0.59 \%$  A supporters is self-eliminated within 4 levels.   Moreover, the number of levels to reach a deterministic outcome decreases with increasing the size group.

The structure of the paper is as follows. In the next section we present the main
frame of the voting model in 
the case of 3-person groups. A critical threshold to deterministic power at the top
is found at $50\%$ of initial support at the bottom level.
The self-elimination of the minority occurs within only a few voting levels.
Section 3 introduces some natural bias in the voting rule with a tip 
to the ruling party in terms of inertia of power. Illustrated in the case of even 4-person groups it yields  an election in favor of the ruling party in case of a local tie 
2A-2B. Such a bias is found to shift the value 
of the critical
threshold to power from $50\%$ to $77\%$ for the opposition. In parallel, it shrinks it from $50\%$ to $23\%$ for the ruling party.
Group size effects are then
analyzed in Section 4.
Analytic formulas are derived in Section 5. Given an A initial support $p_0$,
the number
of voting levels necessary to their self-elimination is calculated.
The results are then turned in a more practical perspective for hierarchical
organizations in Section 6. Some visualization of a numerical
simulation \cite{simu} is shown in Section 7
Extension to 3 competing group is sketched in Section 8.
The last section discusses possible applications of our generic model to describe in part, the 
historical last century collapse of Eastern European communist parties. The results may also shed
a new light on management architectures as well as on alert systems.

%%%%%%%%%%%%%%%%%%%%
\section{The simple case of groups of size 3}

We consider the simplest case with a population divided between 
individuals supporting
either one of two policies A and B. The respective proportions at the 
organization bottom
denoted level-0 are respectively $p_0$ and  $1-p_0$. Each member does 
have an opinion. Neither strategy nor interactions are explicitly included. It means that we do not address the question of why policy A has  a support $p_0$ at the bottom. People do not neither choose their respective voting group to optimize their weight since the groups are formed randomly at each level of the hierarchy.

It is worth to stress that although  we are illustrating the model in terms of voting process, it is quite generic and may apply to many different situations where bottom-up majority rules aggregation are used. The frame may be a political group, a firm, a social organization.

We start studying the case of groups built of 3 persons randomly selected from the population.
It could correspond to home localization or working place. Each group 
then elects
a representative using a local majority rule. Groups with either
3 A or 2 A elect an A. Otherwise it is a B who is elected.
Therefore the probability to have an A elected from the bottom level is,
\begin{equation}
p_1\equiv P_3(p_0)=p_0^3+3 p_0^2 (1-p_0) \ ,
\end{equation}
where $P_3(p_n)$ denotes the voting function, here a simple majority rule.

The same process of group forming is repeated within level-1.
The elected persons (from level-0) form groups which in turn
elect new higher representatives. The new elected persons constitute level-2.
The process can then be repeated again and again.
The probability to have an A elected at level (n+1) from level-n
is,
\begin{equation}
p_{n+1}\equiv P_3(p_n)=p_n^3+3 p_n^2 (1-p_n) \ ,
\end{equation}
where $p_n$ is the proportion of A at level-n.

The analysis of the voting function $P_3(p_n)$ exhibits the existence of
  3 fixed points $p_l=0$, $p_{c,3}=\frac{1}{2}$ and
$p_L=1$. The first one corresponds to not one  A elected.
The last one $p_L=1$ represents the totalitarian situation where only 
A are elected.
Both $p_l$ and $p_L$ are stable fixed points. In contrast $p_{c,3}$ 
is unstable.
It determines indeed the threshold
for flowing towards either full power (with $p_L$) or to total disappearance
(with $p_l$).
Starting from $p_0<\frac{1}{2}$ leads to the first case while
$p_0>\frac{1}{2}$ yields the second. See Figure (2).

\begin{figure}
\centerline{\epsfbox{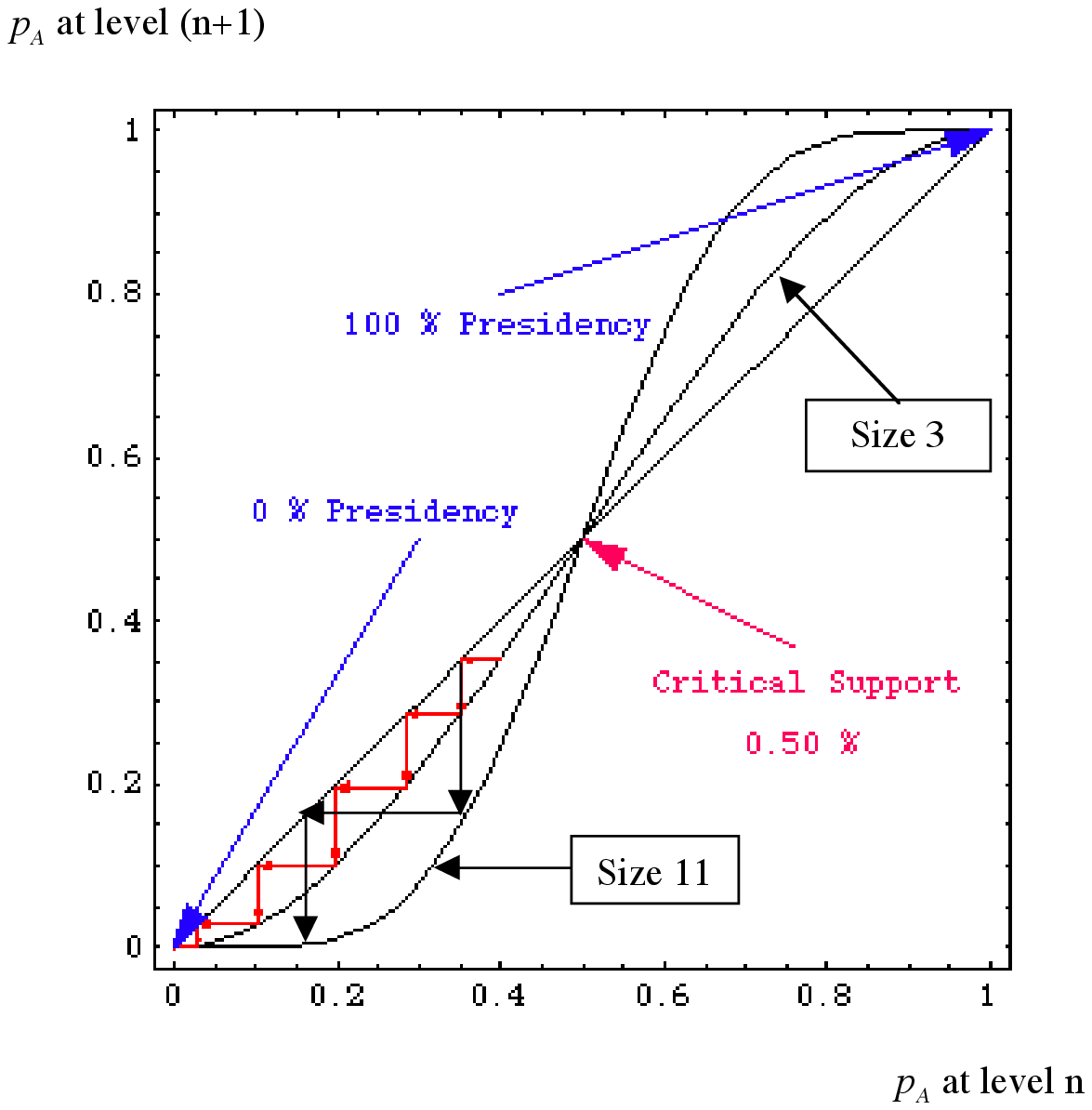}}
\caption{Variation of $p_{n+1}$ as function of $p_{n}$ for groups of 
respective sizes 3 and 11
with both $p_{c,3}=\frac{1}{2}$. Arrows show the direction of the flow.
}
\end{figure}

On this basis we see that majority rule voting produces the self-elimination
of any A proportion as long as the initial support is less than $50\%$
($p_0<\frac{1}{2}$). However this democratic
self-elimination requires a sufficient number of voting levels to be completed.

At this stage the instrumental question is to determine the number of
levels required to ensure full leadership to the initial larger party.
To make sense the level number must be small enough,
most organizations having only a few levels (less than 10).

To illustrate the voting dynamics above, let us calculate
the representativity flow starting, for instance, from $p_0=0.45$.
We get successively $p_1=0.42 $,
$p_2=0.39$, $p_3=0.34$, $p_4=0.26$, $p_5=0.17$,
$p_6=0.08$ down to $p_7=0.02$ and $p_8=0.00$. Within 8 levels $45\%$
of the population
is self-eliminated.

Nevertheless the overall process preserves the democratic character 
of majority rule
voting. It is the bottom leading party (more than $50\%$) which eventually gets
for sure the
full leadership of the organization top level.
It is worth to notice the symmetry of situation with respect to A and 
B parties.
The threshold to full power is the same ($50\%$) for both of them.

%%%%%%%%%%%%%%%%%%%%
\section{The tip to be in power: groups of size 4 }

 From the above analysis, to turn down a top leadership requires to have
more than $50\%$ at the bottom which is a fair constraint.
However from real life situations, to produce a change of leadership
is much more difficult. A simple majority appears often not to be enough.

Many institutions are indeed build to strengthen some
stability. Frequent political changes are not perceived as good
for the organization. On this basis, a tip is often given to the ruling party.
To be in charge, gives some additional power which breaks the symmetry between
the two parties. For instance, giving one additional
vote to the committee president or allowing the president do designate
some committee members.

Using the democratic statement ''to change things, you need a majority"
produces indeed a strong bias in favor of current rulers. To exemplify
it, we consider even size groups. Again restricting to the simplest case,
it means 4 people groups.
Assuming the B are the ruling party, the A
need either 4 or 3 A in a given group to take over. The tie case 2A-2B  votes
for a B (see Figure (3)).

\begin{figure}
\begin{center}
\centerline{\epsfbox{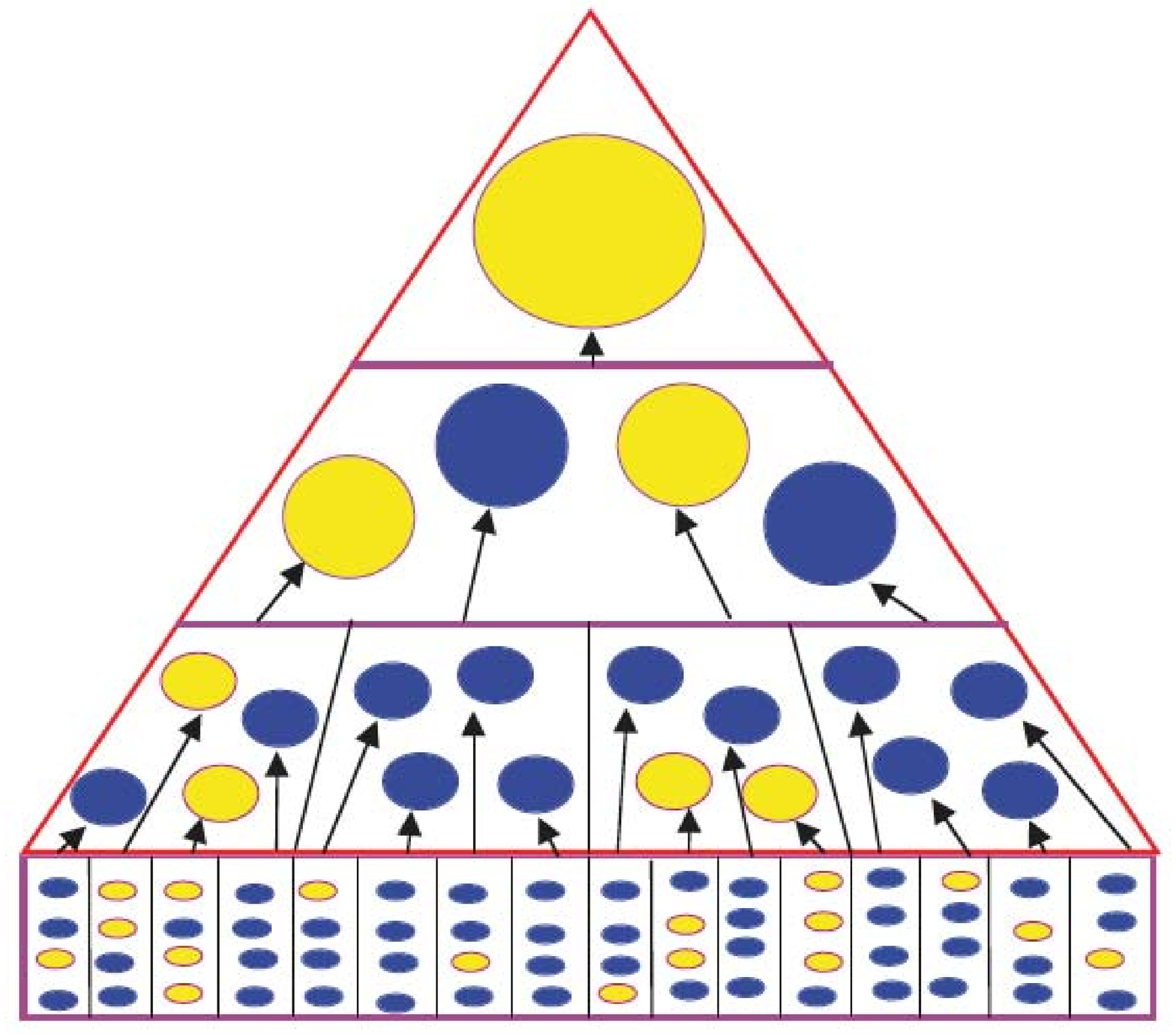}}
\caption{A two-level bottom-up hierarchy with groups of size 4. A tie 
goes to the light color}
\end{center}
\end{figure}

In going from 3 to 4 person size groups, the salient new feature is
indeed the existence
of 2A-2B
configurations for which there exists
no majority. In most social situations no decision implies de facto, no change.
There exists  a bias in favor of the rulers which now makes
the voting function asymmetric. The probability to get an A elected
at level $n+1$ is,
\begin{equation}
p_{n+1}\equiv P_4(p_n)=p_n^4+4 p_n^3 (1-p_n) \ ,
\end{equation}
where $p_n$ is as before the proportion of A elected persons at level-n.
In contrast, for a B to be elected the probability is,
\begin{equation}
1- P_4(p_n)=p_n^4+4 p_n^3 (1-p_n)+2 p_n^2 (1-p_n)^2 \ ,
\end{equation}
where the last term embodies the bias in favor of B. From Eqs (3) and (4) 
the stable
fixed points are still 0 and 1. However, the unstable one is now
drastically shifted to,
\begin{equation}
p_{c,4}=\frac{1+\sqrt{13}}{6} \ ,
\end{equation}
for the A. It makes the A threshold to power at about $77\%$ as shown 
in Figure (4).
Simultaneously
the B threshold to stay in power is about $23\%$ making both 
situations drastically
different. Now, to take over power, the A need to go over $77\%$ of 
initial bottom
support while to stick to power the B only need to keep their support 
above $23\%$.

\begin{figure}
\centerline{\epsfbox{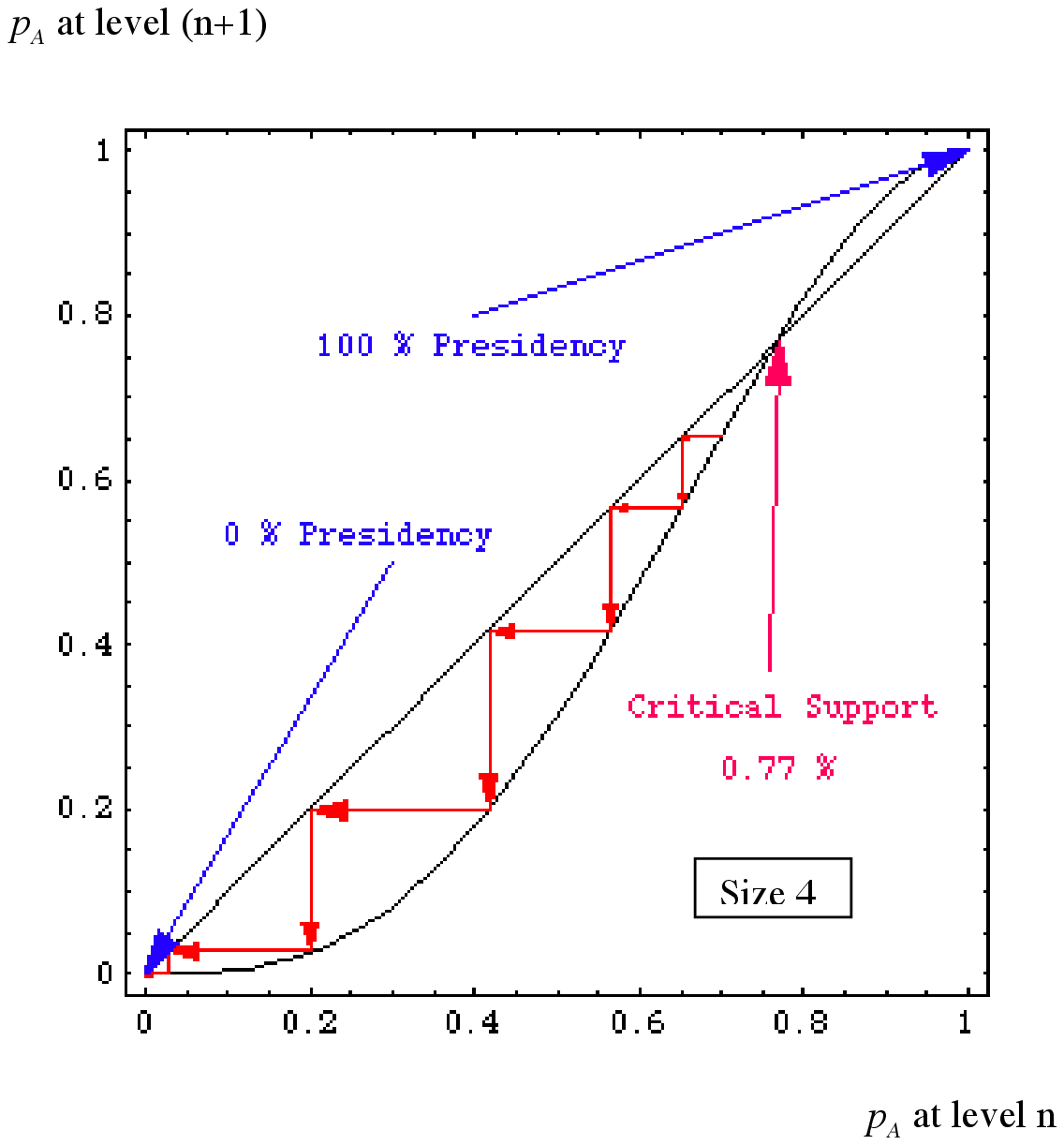}}
\caption{Variation of $p_{n+1}$ as function of $p_{n}$ for groups of respective sizes 4 with $p_{c,3}=\frac{1+\sqrt{13}}{6}\approx 0.77 $. Arrows show the 
direction of the flow.
}
\end{figure}

In addition to the asymmetry the bias makes the
number of levels to democratic self-elimination even smaller than in 
the precedent
case (3-group size). Starting again from $p_0=0.45$ we now get  $p_1=0.24$,
$p_2=0.05$ and $p_3=0.00$. Instead of 8 levels, 3 are enough to make the A 
disappear.

To illustrate how strong this effect is, let us start far above $50\%$ with
for instance $p_0=0.70$. The associated voting dynamics becomes
$p_1=0.66$, $p_2=0.57 $, $p_3=0.42 $, $p_4=0.20 $,
$p_5=0.03 $, and $p_6=0.00$. Within only 6 levels, $70\%$
of a population is thus self-eliminated.

Using an a priori reasonable bias in favor of
the B turns a majority rule democratic voting to a dictatorship outcome.
To get to power the A must pass over $77\%$ of overall support which is
almost out of reach in any normal democratic two party situation.

%%%%%%%%%%%%%%%%%%%%
\section{Larger size groups}

Up to now we have considered very small groups. But many organizations
have larger groups. Extending the above cases to groups of
any size r is indeed straightforward with the
Equations getting a bit more complicated. However the main features remain
unchanged under size changes.

For a r-size group the voting function
$p_{n+1}=P_r(p_{n})$ which accounts for all configurations with a 
majority of A supporters
becomes,
\begin{equation}
P_r(p_n)=\sum_{l=r}^{l=m}\frac{r!}{l!(r-l)!} p_n^l(1+p_n)^{r-l}\ ,
\end{equation}
where $m=\frac{r+1}{2}$ for odd r and $m=\frac{r+1}{2}$ for even r which thus
accounts for the bias in favor of B.

The two stable fixed points $p_l=0$ and $p_L=1$ are unchanged. They are size
independent. In the case of odd sizes, the unstable fixed point is 
also unchanged
with $p_{c,r}=\frac{1}{2}$. On the opposite, for even sizes, the asymmetry
between the threshold values for respectively rulers and non rulers
weakens with increasing sizes.
For the A threshold it is $p_{c,4}=\frac{1+\sqrt{13}}{6}$ for size 4
and it decreases asymptotically
towards
$p_{c,r}=\frac{1}{2}$ for r$ \rightarrow \infty$. It is $65\%$ for size 6,  and stays always larger
than $\frac{1}{2}$ making hard yet to pass the barrier for the opponents.
It is known that in democratic countries a few percent difference between two
candidates is seen as huge.

In parallel increasing group sizes reduces the number of levels 
necessary to get to the stable fixed points as shown in Figure (5). For any application to real social structure, combinations of different sizes should be considered since it is not the same organization which prevails at each level. It is left for future work.

\begin{figure}
\centerline{\epsfbox{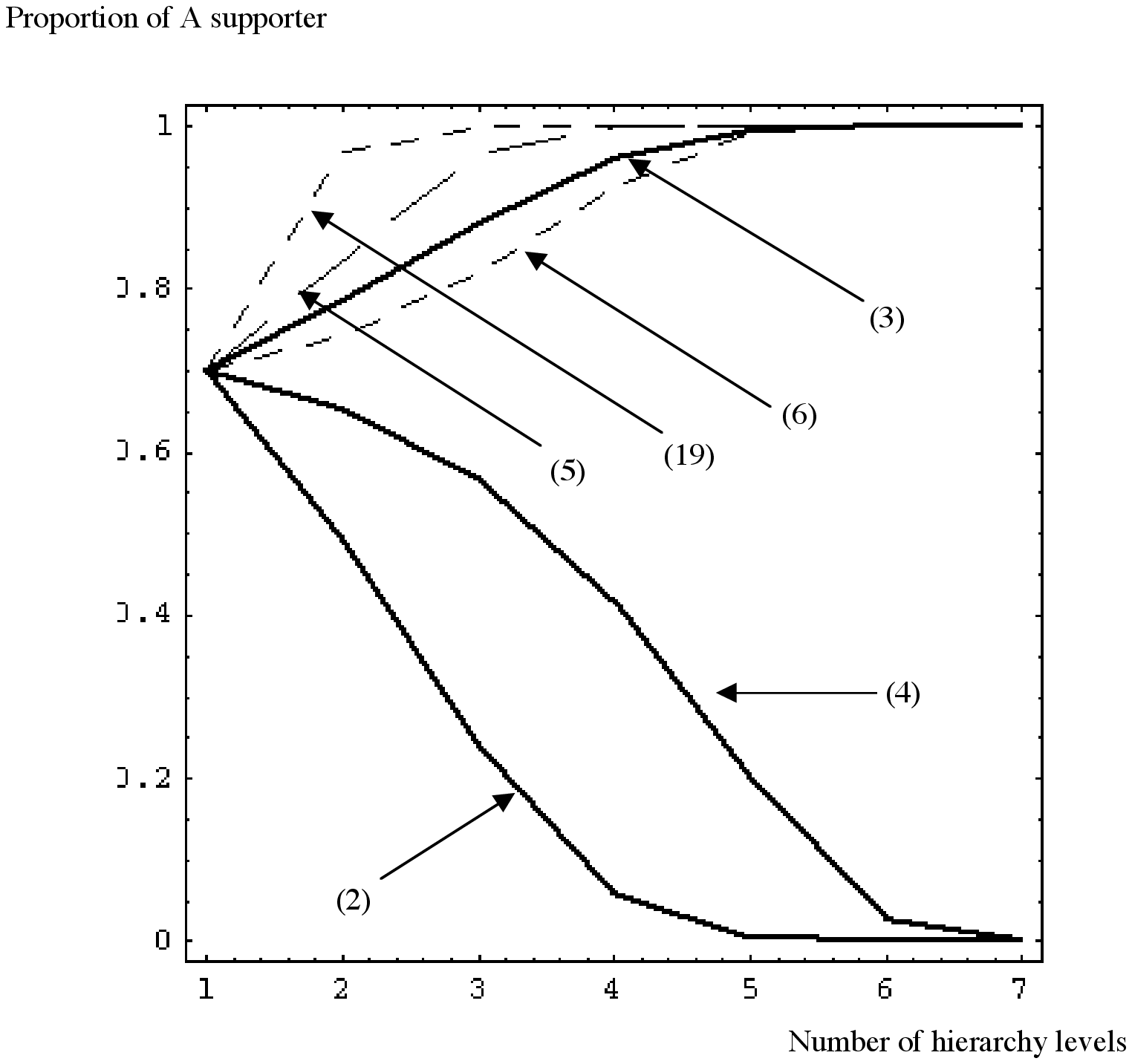}}
\caption{Variation of $p_{n}$ as function of the number of levels for a large
variety of groups sizes. All cases have an initial proportion of 
$70\%$ support for A.
}
\end{figure}

%%%%%%%%%%%%%%%%%%%%
\section{Calculating the critical number of hierarchical levels}

Given an initial support $p_0$ , we want to calculate $p_n$
the corresponding value for A support after n voting levels
as a function of $p_0$. Accordingly
we expand the voting function
$p_n=P_r(p_{n-1})$ around the unstable fixed point $p_{c,r}$,
\begin{equation}
p_n\approx p_{c,r}+(p_{n-1}-p_{c,r}) k_r \ ,
\end{equation}
where $k_r \equiv \frac{dP_r(p_n)}{dp_n}|_{p_{c,r}}$ with
$P_r(p_{c,r})=p_{c,r}$. Rewriting
the last Equation as
\begin{equation}
p_n-p_{c,r}\approx (p_{n-1}-p_{c,r}) k_r \ ,
\end{equation}
we can then iterate the process to get,
\begin{equation}
p_n-p_{c,r}\approx (p_0-p_{c,r}) k_r^n \ ,
\end{equation}
from which we get,
\begin{equation}
p_n\approx p_{c,r}+(p_0-p_{c,r}) k_r^n \ .
\end{equation}

 From Eq. (10) two different critical numbers of
levels $n_c^l$
and $n_c^L$ can be obtained. the first one corresponds to $p_{n_c^l}=0$
and the second one to $p_{n_c^L}=1$.
Putting $p_n=p_{n_c^l}=0$ in Eq. (10) gives,
\begin{equation}
n_c^l\approx \frac{1}{\ln k_r}  \ln\frac{ p_{c,r}}{ p_{c,r}-p_0}  \ ,
\end{equation}
which is defined only for $p_0<p_{c,r}$ showing that only below 
$p_{c,r}$ can the
proportion decrease to zero. On the other hand putting $p_n=p_{n_c^L}=1$
in the same Eq. (10) gives,
\begin{equation}
n_c^L\approx \frac{1}{\ln k_r}  \ln\frac{ p_{c,r}-1}{ p_{c,r}-p_0}  \ ,
\end{equation}
which is now defined only for $p_0>p_{c,r}$ since $p_{c,r}<1$,
showing that only above $p_{c,r}$ can the
proportion increase to one.

Though above expansions are a priori valid only
in the vicinity of $p_{c,r}$, they turn out to be rather good estimates
even in getting close to the two stable fixed points 0 and 1.  While taking
the integer part of Eqs. (11) and (12) the rounding is
always to the larger value.

%%%%%%%%%%%%%%%%%%%%
\section{Inversing the problem}

However once organizations are set, they are usually not modified.
Therefore within  a given organization the number
of hierarchical levels is a fixed quantity.
Along this line, to make a practical use of above
analysis the question of ``How many level are needed to
eliminate a party?" must be turned to,
\begin{center}
'`Given n levels what is the minimum overall support\\
to get full power for sure?".
\end{center}
Or alternatively, for the ruling group,
\begin{center}
'`Given n levels what is the critical overall support of the competing party
below which it always self-eliminates totally?",
\end{center}
which means no worry for the ruling party about its current policy.

To implement this operative question, we rewrite Eq. (10) as,
\begin{equation}
p_0=p_{c,r}+(p_n-p_{c,r}) k_r^{-n} \ .
\end{equation}
It yields two critical thresholds.
First one is the disappearance threshold $p_{l,r}^n$ which gives the value
of support under which the A disappear for sure at the top level
of the n-level hierarchy.
It is given by Eq. (13) with $p_n=0$,
\begin{equation}
p_{l,r}^n=p_{c,r}(1-k_r^{-n}) \ .
\end{equation}
In parallel $p_n=1$ gives the second threshold $p_{L,r}^n$ above which
the A get full and total power. From Eq. (13),
\begin{equation}
p_{L,r}^n=p_{l,r}^n+k_r^{-n}\ .
\end{equation}
There exists now a new regime for $p_{l,r}^n<p_0<p_{L,r}^n$.
In there, A neither disappears totally nor gets to full power.

It is a coexistence region where some democracy is prevailing since results
of the election process are only probabilistic. No party is sure of winning
making alternate leadership a reality. However as seen from Eq. (15), this
democratic region shrinks as a power law $k_r^{-n}$ of
the number n of hierarchical levels.
A small number of levels puts higher the threshold to a total
reversal of power but simultaneously lowers the threshold for non-existence.

Above formulas are approximates since we have neglected
corrections in the vicinity of the stable fixed points. However we found they
give the right quantitative  values by adding 1 to $p_{l,r}^n$ and 2 
to $p_{L,r}^n$  \cite{math-psy}.

To get a practical feeling of what Eqs. (14) and (15) mean, let us illustrate
the case $r=4$ where $k =1.64$ and $p_{c,4}=\frac{1+\sqrt{13}}{6}$.
Considering 3, 4, 5, 6 and 7 level organizations, $p_{l,r}^n$ is equal to
respectively $0.59$, $0.66$, $0.70$, $0.73$ and $0.74$. In parallel $p_{L,r}^n$
equals $0.82$, $0.80$, $0.79$, $0.78$ and $0.78$.
These series emphasize drastically the totalitarian character of the 
voting process.

\section{Results from a simulation: visualizing the dynamics}

To exhibit the strength of the phenomena, we show some snapshots of a
numerical simulation \cite{simu}. The two A and B parties are represented
respectively in white and black squares,
with the bias in favor of the black ones, i. e., a tie 2-2 votes for a black
square. A structure with 8 levels is shown. We can see on each 
picture, how a huge
white square majority is self-eliminated. Written percentages are for the white
representation at each level. The ``Time" and ``Generations"
indicators should be discarded.

Figure (6) shows a case where $52.17\%$ of the people at the bottom are in support of the  B (white), thus a bit over the expected $50\%$ democratic threshold to take over. However,  3 levels higher in the hierarchy no white squares appear. The bottom majority has self-evaporated.
Figure (7) shows the same population with now a substantial increase in B(white) support with a proportion of $68.62\%$, now rather  far  than $50\%$. And yet, after 4 levels no more white (B) square is found.
Figure (8) shows the situation where the B (white) support has climbed up to the huge value of $76.07\%$. But again 6 levels higher a A (black) is elected with certainty though its bottom support is as low as support (black) $23.03\%$.
Finally Figure (9) shows an additional   $O.08\%$ very weak increase in B (white) support, 
putting the actual support at  $77.05\%$, which in turn prompts an elected B (white) president.

Some insight about  the often observed blindness of leaderships towards huge and drastic calls for changement from the botton level of an organization can be gained from the simulation. Indeed it shows how and why a president who would get some information about the possible disagreement with a policy cannot account for it since it has to check it from what reaches it from the bottom of the hierarchy. As seen from Figure (7) while the opposition is at already at a pick of $68.62\%$ the president gets $100\%$ of totally satisfied votes from the two levels below it and overwhelming satisfied in the two more lower levels, so why should it make any policy change ?

\begin{figure}
\begin{center}
\centerline{\epsfbox{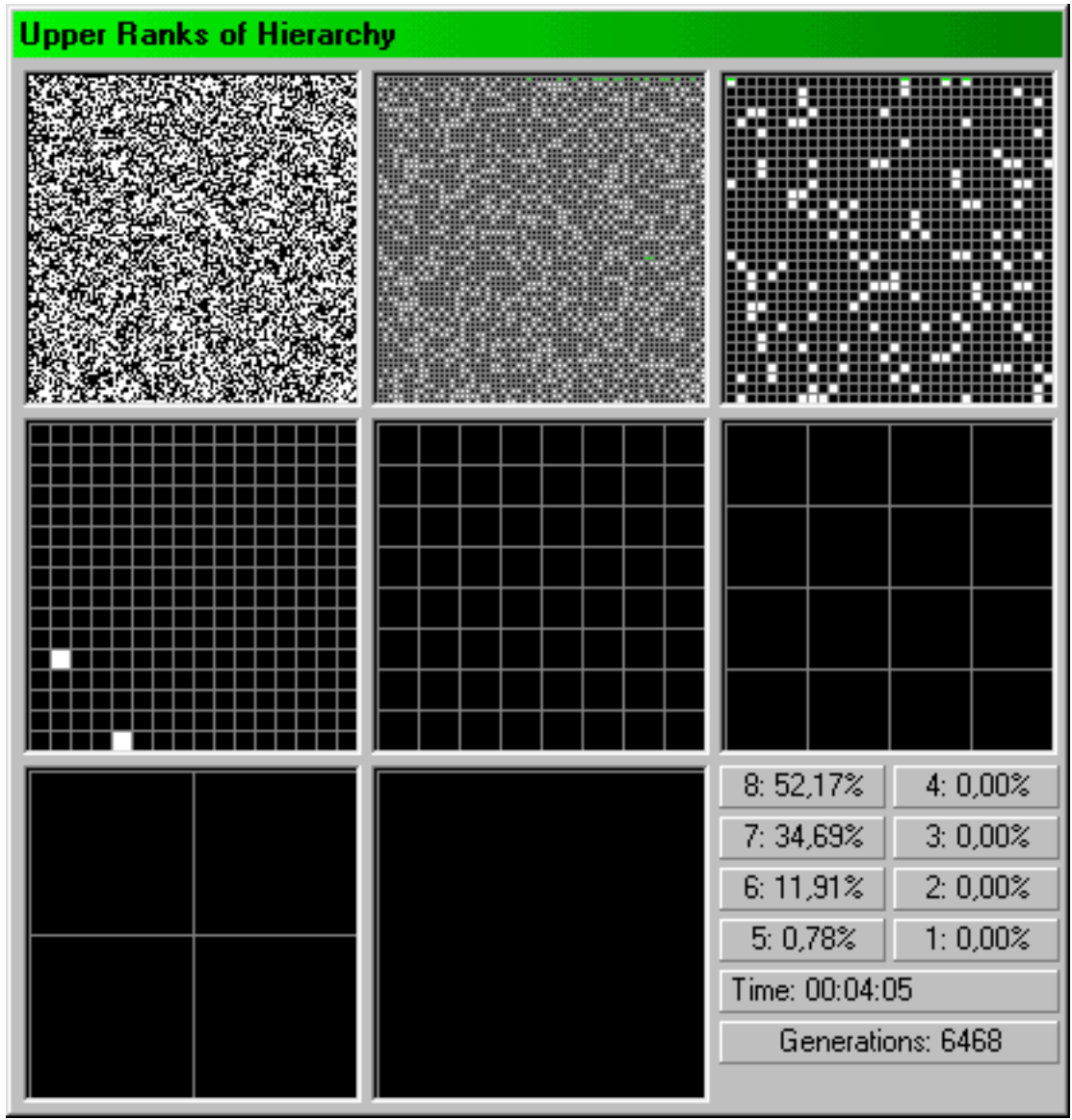}}
\caption{A 8 level hierarchy for even groups of 4 persons.
The two A and B parties are represented respectively in white and black
with the bias in favor of the black squares, i. e., a tie 2-2 votes for a black
square. Written percentages are for the white
representation at each level. The ``Time" and ``Generations"
indicators should be discarded. The initial white support is $52.17\%$.
}
\end{center}
\end{figure}

\begin{figure}
\begin{center}
\centerline{\epsfbox{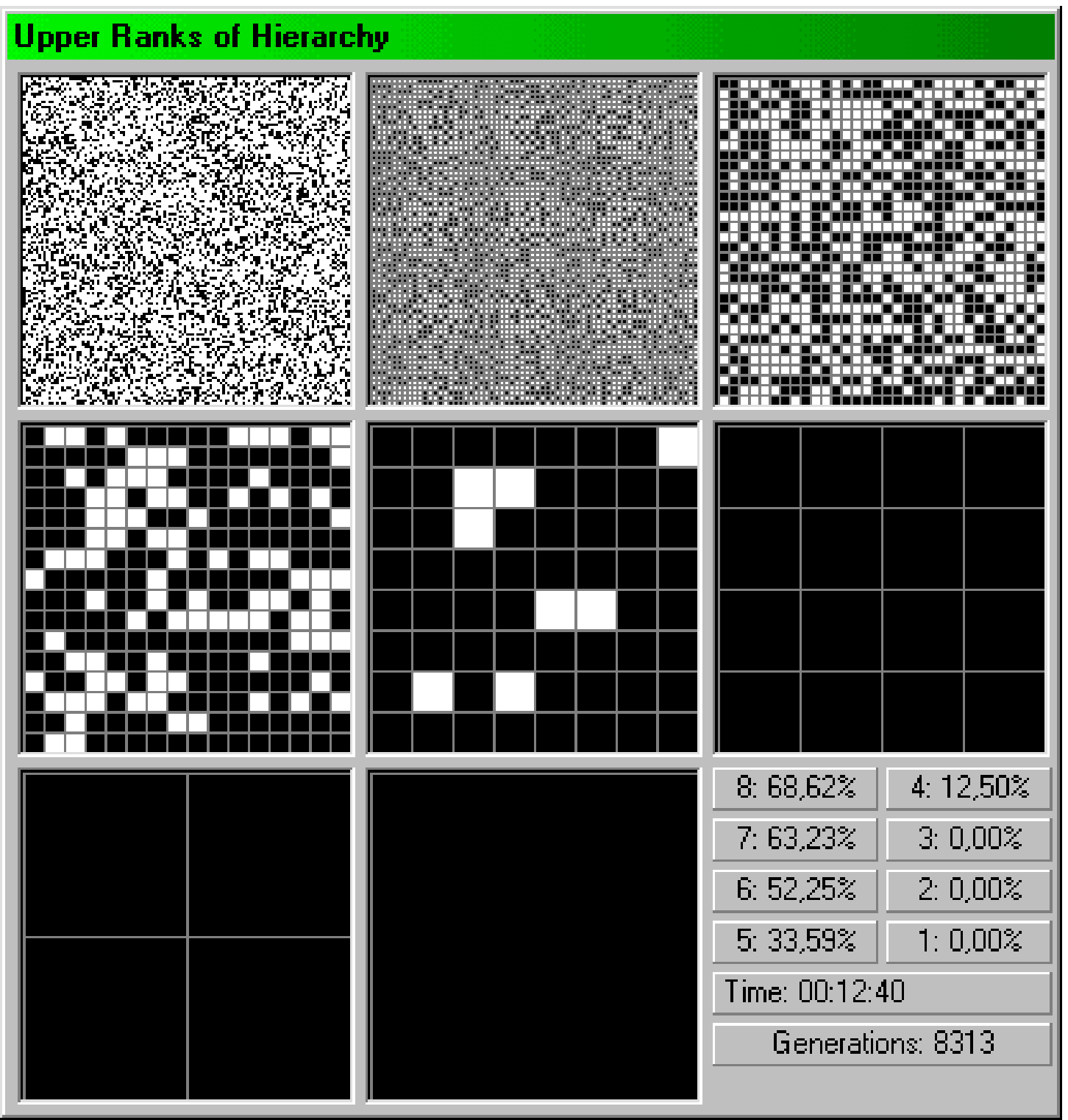}}
\caption{The same as Figure 6 with an initial white support of
$68.62\%$. The presidency stays black.
}
\end{center}
\end{figure}

\begin{figure}
\begin{center}
\centerline{\epsfbox{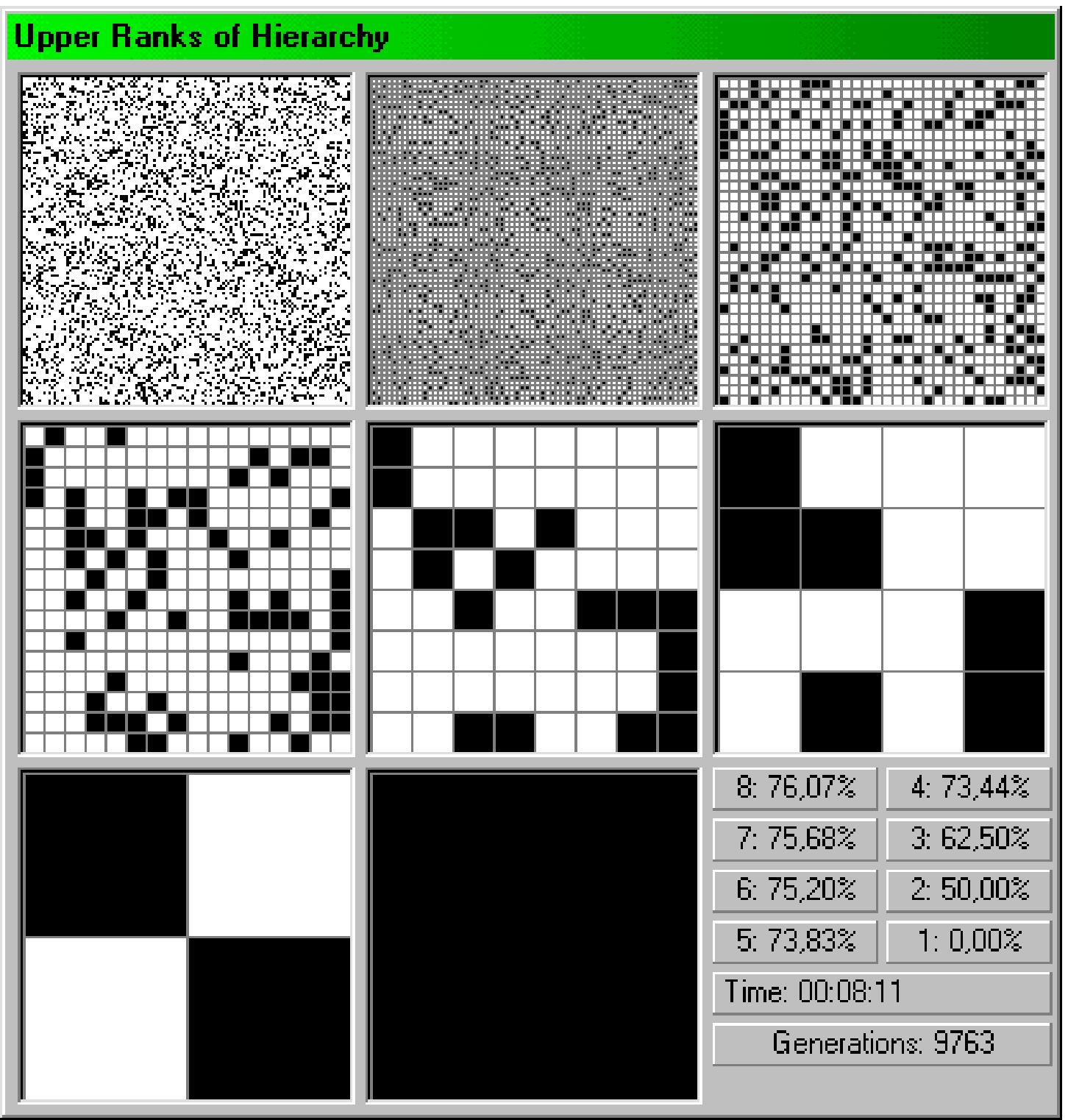}}
\caption{The same as Figure 6 with an initial white support of
$76.07\%$. The presidency stays black.}
\end{center}
\end{figure}

\begin{figure}
\begin{center}
\centerline{\epsfbox{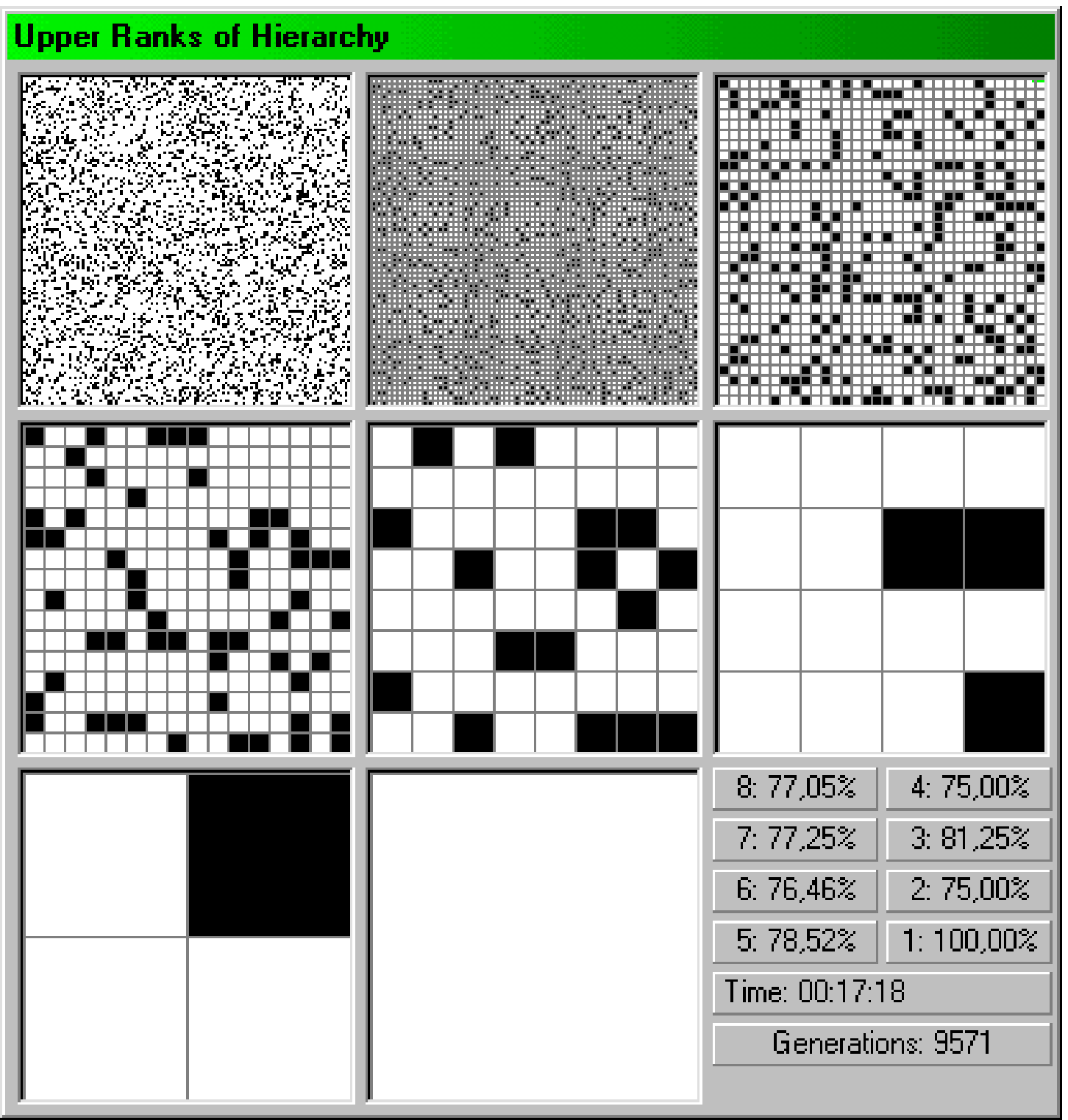}}
\caption{The same as Figure 6 with an initial white support of
$77.05\%$. The presidency finally turned white.
}
\end{center}
\end{figure}

\section{The case of 3 competing groups}

Up to now we have treated very simple cases to single out some main 
trends produced
by democratic voting while repeated over several hierarchical levels.
In particular
the existence of critical thresholds was found to be instrumental
in predicting the voting outcome. Moreover these thresholds
are not necessarily symmetric with respect to both competing parties.
In the  4-group
case it can be $0.77\%$ for the opposition and $0.23\%$ for the rulers.
Such asymmetries are indeed always present one way or another.

In addition, most real situations involve more than
two competing groups. Let us consider for instance the case of three competing
groups A, B and C. Assuming a 3-group case, the ABC configuration is 
now unsolved
since it has no majority the same way as in the case of the two party 
configuration
AABB with groups of size 4.
For the AABB case we made the bias in favor of the ruling group.
For multi-group competition typically the bias is decided  in each configuration and from the various parties agreement.

Usually the two largest parties, say A and B are hostile to each other,
while
the smallest one C would compromise with either one of them.
Along this line, the ABC configuration elects a C in a coalition with 
either A or B.
In such a case, we need 2A or 2B to elect respectively an A or a B.
Otherwise a C is elected.

Therefore the elective functions for A and B are the same as for the A in
the two party  3-group model. It means that the critical threshold to full
power to A and B
is $50\%$. In other words for initial support to both A and B,
less than $50\%$ the C  get full
power. The required number of levels is obtained from the 3-group formula
(Equations (11) and (13) with $r=3$).

Let us illustrate what happens for initial support, for instance of $39\%$
for both A and B. The C are thus left with $22\%$. We will have the series
for respectively A, B and C: $34\%$, $34\%$, $32\%$ at  the first level;
$26\%$, $26\%$, $48\%$ at  the second level;
$17\%$, $17\%$, $66\%$ at  the third level;
$8\%$, $8\%$, $84\%$ at  the forth level;
$2\%$, $02\%$, $96\%$ at  the fifth level;
and $0\%$, $0\%$, $100\%$ at  the sixth level giving total
power to the C minority within only 6 levels.

It is possible to generalize the present approach to as many parties as wanted.
The analysis becomes much more heavy but the mean features of voting flows
towards fixed point are preserved. Moreover power will go even more 
rarely to the
largest groups as seen with the above ABC case \cite{3choices}.

\section{Conclusion}

To conclude we emphasize on the very generic character of our model which allows to consider many different applications. One could be related to last century's
auto-collapse of eastern european communist parties.
Up to this historical and drastic
event, communist parties seemed eternal. Once they
collapsed many explanations were based on some hierarchical opportunistic change within the
various organizations, with in addition for the non russian ones,
the end of the soviet army threat. Maybe our hierachical model can provide some different new insight at such a unique event.

Communist organizations are indeed based, at least in principle,
on the concept of democratic centralism which is a tree-like hierarchy.
Suppose that the critical threshold to power was of the order of $77\%$ like
in our size 4 case.
We could then consider that the internal opposition to the orthodox leadership
did grow a lot and massively over several decades to eventually reach and pass
the critical threshold. Once the threshold was passed,
then we had the associated sudden rise to the leadership levels of the internal opposition.
Therefore, what looked  at a sudden and punctual  collapse of Eastern European
communist parties could have been indeed the result
of a very long and solid phenomenon inside the communist parties.
Such an explanation does not oppose to the very many additional
features which were instrumental in these collapses. It only  singles out some trend within 
the internal mechanism of these organizations which makes them extremely stable.

The results may also shed a new light on management architectures as well as on alert systems. However in order to avoid any confusion between political issues and the management of firms and organizations, it should be noticed that if there exist some similarities between a government as an organization while some constraints are inevitable in management itself, both realities are quite different in nature and objectives.

%%%%%%%%%%%%%%%%%

%%%%%%%%%%%%%%%%%%%

%%%%%%%%%%%%%%%%%

\newpage
\begin{thebibliography}{99}

\bibitem {general} Black D. (1972), "  The theory of Commitee and 
Elections ", Cambridge University Press, 1972,
pp. 183-185.

\bibitem {review-voting} http://www.wikipedia.org/wiki/Voting$\_$system

\bibitem {math-psy} S. Galam,
``Majority rule, hierarchical structures and democratic
totalitarism: a statistical approach", J. of Math. Psychology 
\textbf{30}, 426-434
(1986)

\bibitem {grano} M. Granovetter, ``Threshold models of collective 
behavior", American
Journal of Sociology 83, 1420-1443 (1978)

\bibitem {schelling} T. C. Schelling Micromotives and Macrobehavior, New York,
Norton and Co. (1978)

\bibitem {lemonde} S. Galam Les r\'eformes sont-elles impossibles ?,
Le Monde/ 28 mars/ 18-19 (2000)

\bibitem {status} R. D. Friedman and M. Friedman, The Tyranny of the 
Status Quo,
Harcourt Brace Company (1984)

\bibitem {simu} S. Galam and S. Wonczak, Dictatorship from majority 
rule voting,
Eur. Phys. J. B  18, 183-186 (2000)

\bibitem {3choices} S. Gekle, Luca Peliti and S. Galam, Opinion dynamics in a three-choice system,
Eur. Phys. J. B in Press (2005)

\end{thebibliography}
\end{document}